\documentclass[letterpaper,prb,twocolumn,preprintnumbers]{revtex4}
\usepackage{graphicx,bm}
\begin{document}
\newcommand{\mybm}[1]{\mbox{\boldmath$#1$}}
\newcommand{\mysw}[1]{\scriptscriptstyle #1}

\title{Potential Flow of Renormalization Group in Quasi-One-Dimensional Systems}

\author{Wei Chen}
\affiliation{Department of Physics, University of Florida, Gainesville, FL 32611}
\author{Ming-Shyang Chang}
\affiliation{Physics Department, Boston University, Boston, MA02215}
\author{Hsiu-Hau Lin}
\altaffiliation[]{{\tt hsiuhau@phys.nthu.edu.tw}, on leave from Department of Physics, National Tsing-Hua University, Hsinchu 300, Taiwan}
\affiliation{Kavli Institute for Theoretical Physics, University of California, Santa Barbara, CA 93106}
\affiliation{Physics Division, National Center for Theoretical
Sciences, Hsinchu 300, Taiwan}
\author{Darwin Chang}
\author{Chung-Yu Mou}
\affiliation{Department of Physics, National Tsing-Hua University, Hsinchu 300, Taiwan}
\affiliation{Physics Division, National Center for Theoretical
Sciences, Hsinchu 300, Taiwan}
\date{\today}

\begin{abstract}
We address the issue why the phase diagrams for quasi-one-dimensional systems are rather simple, while the renormalization group equations behind the scene are non-linear and messy looking. The puzzle is answered in two steps -- we first demonstrate that the complicated coupled flow equations are simply described by a potential $V(h_i)$, in an appropriate basis for the interaction couplings $h_i$. The renormalization-group potential is explicitly constructed by introducing the Majorana fermion representation. The existence of the potential prevents chaotic behaviors and other exotic possibilities such as limit cycles. Once the potential is obtained, the ultimate fate of the flows are described by a special set of fixed-ray solutions and the phase diagram is determined by Abelian bosonization.  Extension to strong coupling regime and comparison with the Zamolodchikov c-theorem are discussed at the end.
\end{abstract}
\maketitle

\section{Introduction}
Quasi-one-dimensional (Q1D) systems have attracted extensive attentions
from both experimental and theoretical aspects\cite{Dagotto96}. 
Due to the low
dimensionality, strong quantum fluctuations often give rise to
surprising behaviors, which are rather different from our
intuitions built in higher dimensions. It is then exciting to
explore various exotic phenomena, such as spin-charge separation\cite{Kim96},
unconventional electron pairing\cite{Noack94,Balents96} and so on, in the correlated Q1D systems\cite{Fabrizio93,Khveshchenko96,Schulz96,Lin97,Lin98,Ledermann00}. Note that, in addition to the ladder compounds, carbon nanotubes, nanoribbons and quantum wires, after integrating out fluctuations at higher energy, are also described by the Q1D theory. Therefore, not only posting a challenging task for academic curiosity, the understanding of the Q1D systems now becomes
crucial as the technology advances to the extremely small
nanometer scale.

Although our understanding of the strictly one-dimensional systems is greatly
benefitted from exact solutions, it is often found that the Q1D
systems are not soluble analytically. Furthermore, since the hopping along the transverse direction is relevant, the physics phenomena for Q1D systems can be dramatically different from the strictly 1D systems, such as electrons with mutual repulsive interactions are found to form unconventional ``d-wave'' Cooper pairs in the ladder systems. Therefore, the most reliable approach to clarify the competitions among various ground states is the renomalization group (RG) analysis. Since the
number of allowed interactions are large, the derivation of the
flow equations for all couplings under RG transformations becomes
formidable. On the other hand, both numerical and analytical
approaches seem to indicate rather simple phase diagrams for
generic Q1D systems. 

The simplicity of the phase diagram can be partially understood by the widely used scaling Ansatz of the couplings in weak coupling\cite{Balents96,Lin97,Lin98,Lecheminant04},
\begin{eqnarray}
g_{i}(l) \simeq \frac{G_{i}}{(l_{d}-l)} \ll 1,
\label{Scaling}
\end{eqnarray}
where $G_{i}$ are order one constants satisfying the non-linear algebraic constraints (discussed later) and $l_{d}$ is the divergent length scale where the flows enter the strong coupling regime. The Ansatz was motivated by the numerical observation that the ratios of renormalized couplings reach constant, as long as the bare interaction strength is weak enough. However, it is still puzzling why the phase diagrams,
generated by many coupled non-linear flow equations, do
not reflect the same level of complexity. In fact, for a complicated system with many couplings, the coupled non-linear differential equations are likely to produce chaotic
flows generically. Even if the flow is not chaotic, it might as
well rest on limit cycles. This peculiar possibility for quantum
systems was addressed by Wilson and collaborators in a recent
paper\cite{Glazek02}. 

So the question remains: Why are the phase diagram and the RG
flows so simple (without chaotic flow or limit cycle) in Q1D
systems? We found the question can be answered by combining the weak-coupling RG analysis together with the non-perturbative Abelian bosonization technique. Note that the phases of the correlated ground state often lie in the strong coupling, so weak-coupling RG alone can not pin down the phase diagram. It is the powerful combination of the perturbative RG analysis and the non-perturbative bosonization technique which deliver the desired answer here.

Let us start with the weak-coupling RG. In the low-energy limit, the Q1D systems involve $N_f$ flavors of interacting fermions with {\em different} Fermi velocities, where
$N_f$ is the number of conducting bands. Although constrained by various symmetries, the number of allowed interactions is tremendously large as $N_f$ increases. In general, the RG equations to the lowest order are already very complicated
- not to mention solving them analytically.

However, quite to our surprises, we found that, at one-loop level,
the RG flows cab be derived from a potential, i.e. the coupled non-linear flow equations can be cast into this elegant
form by an appropriate choice of coupling basis,
\begin{equation}
\frac{dh_i}{dl} = - \frac{\partial V(h_{j})}{\partial h_i},
\label{PotentialFlow}
\end{equation}
where $V(h_{j})$ is the RG potential. 
We emphasize that this is only possible after a unique transformation
of the couplings, $h_i(l) = L_{ij} g_j(l)$ (up to a trivial overall
factor for all couplings), where $L_{ij}$ is some constant matrix.
The existence of the potential, which requires the coefficients in the RG equations to 
satisfy special constraints, also provides a self-consistent check on the RG equations derived by other approaches.

The flows of Eq.~(\ref{PotentialFlow})
in the multi-dimensional coupling space can be viewed as the
trajectories of a strongly overdamped particle moving in a
conservative potential $V(h_i)$. Note that the change of the potential
$V(h_i)$ along the trajectory is always decreasing,
\begin{equation}
\frac{dV}{dl} = \frac{\partial V}{\partial h_i} \frac{dh_i}{dl} =
- \left(\frac{dh_i}{dl}\right)^2 \leq 0,
\end{equation}
where summation over the index $i$ is implicitly implied.
Therefore, it is obvious that the function $V(h_{i})$ never increases
along the trajectory and is only stationary at the fixed points
where $dh_i/dl=0$. Thus, the RG flows have a simple geometric
interpretation as the trajectory of an overdamped particle searching for potential
minimum in the multi-dimensional coupling space.

This simple geometric picture rules out the possibilities of chaos and
the exotic limit cycles in Q1D systems. 
The ultimate fate of the flows would either rest on
the fixed points or follow along the  ``valleys/ridges" of the
potential profile\cite{Lin00,Konik02} to strong coupling. Since there is only one trivial fixed point at one-loop order, most of the time, the flows run away from the non-interacting fixed point. Starting from weak enough bare couplings, the ultimate fate of the flows is dictated by the asymptotes of the ``valleys/ridges" of the potential profile. It provides the natural explanation why the ratios of the renormalized couplings reach constant in numerics. That is to say, the existence of RG potential implies that the ultimate fate of RG flows must take the scaling form described in Eq.~\ref{Scaling}. Detail properties of these asymptotes, referred as fixed rays, will be discussed in later section. 

Since the ultimate fate of RG flows is described by the simple Ansatz in Eq.~\ref{Scaling}, the specific ratios of couplings simplify the effective Hamiltonian a lot. Making use of the Abelian bosonization, one can determine which sector acquires a gap, triggered from the weak-coupling instability. The phase of the ground state is then determined by watching which fixed ray (asymptote) the flows go closer to. Because there are only limited solutions of the fixed rays, the phase diagram is rather simple. Therefore, by combining the powerful techniques of weak-coupling RG and Abelian bosonization, we pin down the reason behind the simple-looking phase diagram out of the messy non-linear flow equations.

In fact, the combination of weak-coupling RG and Abelian bosonization goes beyond the usual mean-field analysis and is crucially important when there are more than one competing orders\cite{Lauchli04}. For instance, Lauchli, Honercamp and Rice recently studied the so-called ``d-Mott" phase in one and two dimensions,  where antiferromagnetic, stagger-flux and d-wave pairing fluctuations compete with each other simultaneously. The conclusion drawn from the numerical density-matrix RG in strong coupling agrees rather well with predictions made from one-loop analysis in weak coupling. This lends support to the powerful combination of weak-coupling RG and Abelian bosonization approach for strongly correlated systems.

Since the method of bosonizing the fixed rays is already developed in previous papers\cite{Fabrizio93,Balents96,Lin97,Lin98}, we would concentrate on the novel existence of RG potential in this paper. In particular, we would construct the RG potential explicitly. The details of the bosonization and numerical results will be deferred for future publication. The remaining part of the paper is organized in the following: In Sec. II, the criterion for potential flows and the notation of fixed rays are explained. In Sec. III, we write down the effective Hamiltonian for generic Q1D systems and elaborate all possible interaction vertices in weak coupling. In Sec. IV, the Hamiltonian is re-written in terms of Majorana fermions and the RG potential is explicitly constructed. Finally, we close the paper by the section on discussions and the summary of the main results.

\section{Criterion for Potential Flows}

To prove the existence of the RG potential, it is helpful to study the general feature of one-loop RG equations of the Q1D systems first. In weak
coupling, the most relevant interactions are the marginal four-fermion interactions, described by a set of dimensionless couplings $g_i$.
The RG transformation to the one-loop order is described by a set of coupled non-linear first-order differential equations
\begin{equation}
\frac{dg_i}{dl} = M^{jk}_{i} g_j g_k \equiv F_i,
\label{OneLoop}
\end{equation}
where the coefficients $M^{jk}_{i} = M^{kj}_{i}$ are symmetrical
by construction. These constant tensors $M^{jk}_{i}$ completely
determine the RG flows.

The solution for Eq.~(\ref{OneLoop}) can be viewed as
the trajectory of a strongly overdamped particle under the
influence of the external force $F_i$ in the multi-dimensional
coupling space. Naively, one might rush to the conclusion that the
conditions for the existence of a potential requires are,
\begin{equation}
\frac{\partial F_i}{\partial g_j} - \frac{\partial F_j}{\partial g_i} =0, \hspace{4mm}
\rightarrow \hspace{4mm} M_{i}^{jk} = M_{j}^{ik},
\end{equation}
which implies that the tensor $M_{i}^{jk}$is totally symmetric. It
is straightforward to check that the RG equations for the Q1D
systems {\em do not} satisfy this criterion \cite{Lin97,Ledermann00,Chang04}.

However, under general linear transformations of the couplings $h_i(l) = L_{ij} g_{j}(l)$, the coefficients $M_{i}^{jk}$ transform into a new set of coefficients $N_{i}^{jk}$, which may become symmetric. For convenience, we introduce a set of matrices $[\bm{M}(k)]_{ij}\equiv M_{i}^{jk}$ to represent the coefficients. The symmetric criterion for $N_{i}^{jk}=N_{j}^{ik}$ requires the existence of a constant matrix $\bm{L}$ which satisfies the following constraints (for all $k$!), 
\begin{equation}
\bm{M}(k)^T = (\bm{L}^{T} \bm{L}) \bm{M}(k)( \bm{L}^{T} \bm{L})^{-1},
\end{equation}
where superscript $T$ means transpose. In general, there is no guarantee why the strongly over-determined constraints would allow a solution for $\bm{L}$. In fact, it is a nontrivial task to just prove/disprove whether the desired linear transformation $\bm{L}$ exists. Surprisingly, for the Q1D systems, the hunt for the solution greatly simplifies if we formulate the problem in terms of Majorana fermions. The desired transformation becomes diagonal, $L_{ij} = r_{i} \delta_{ij}$, where $r_i$ is a set of rescaling factors and leads to the totally symmetric coefficients,
\begin{equation}
N^{jk}_{i} = \left(\frac{r_i}{r_j r_k}\right) M^{jk}_{i}.
\end{equation}
So the search for the potential is now nailed down to find a set of rescaling factors $r_i$ in the Majorana representation. In later section, we demonstrate how to construct the RG potential explicitly in the Majorana representation. In fact, one can also construct the potential $V(h_i)$ directly from the RG equations for doped\cite{Lin97} and half-filled\cite{Ledermann00,Chang04} Q1D systems. Both approaches lead the the same result and the detail work will be described elsewhere.

Before we leave this section, it is important to discuss a special set of analytic solution of Eq.~\ref{OneLoop}, which is closely related to the scaling Ansatz in Eq.~\ref{Scaling}. Suppose the initial values of the couplings are $ g_{i}(0) =  G_{i} g(0) $, where $g(0)=U \ll 1$ and $G_{i}$ are order-one constants satisfying the non-linear algebraic constraint,
\begin{equation}
G_{i} = M^{jk}_{i} G_{j} G_{k}.
\label{Algebraic}	
\end{equation}
It is straightforward to show that the ratios between couplings 
remain the same and the complicated equations reduce to single 
one,
\begin{eqnarray}
\frac{dg}{dl} = g^{2}.
\end{eqnarray}
For repulsive interaction $U>0$, the above equation can be solved easily $g(l) = 1/(l_{d}-l)$, where the divergent length scale $l_{d} = 1/U$. Note that this implies the ratios of different couplings remain fixed in the RG flows,
\begin{eqnarray}
g_{i}(l) = \frac{G_{i}}{l_{d}-l}.
\end{eqnarray}
These special analytic solutions are referred as ``fixed rays'' because the ratios of the renormalized couplings remain fixed along the flows. One immediately notices that these special set of solutions are nothing but the peculiar Ansatz found in the numerics. As explained in the introduction, if the RG potential exists, these fixed rays are the asymptotes of the ``valleys/ridges'' of the potential profile and capture the ultimate fate of RG flows completely.

\section{Quasi-1D Ladder}

Since the readers might not be familiar with Majorana representation, it is worthwhile to write down the Hamiltonian in terms of the familiar field operators for electrons first. To construct the RG potential, we first write down the general interacting Hamiltonian in weak coupling. To be concrete, we would take the $N$-leg ladder as the example. However, the general framework developed here can be applied to more general Q1D systems. In weak coupling, it is natural to work on the band structure first. Suppose the chemical potential cuts through $N_{f}$ flavors of bands at different Fermi momenta $k_{\mysw{F}i}$ with velocities $v_{i}$. In the low-energy limit, the electron operator can be decomposed into chiral fields,
\begin{equation}
\psi_{i\alpha}(x) \sim 
\psi^{}_{\mysw{R}i\alpha}(x) \: e^{ik_{\mysw{F}i}x}
+\psi_{\mysw{L}i\alpha}(x)\: e^{-i k_{\mysw{F}i}x},
\label{chiral}
\end{equation}
where $\alpha$ is the spin index, and $i$ stands for the band index. The effective Hamiltonian density is simply a collection of $N_{f}$ flavors of Dirac fermions,
\begin{eqnarray}
{\cal H}_{0} = \psi^{\dag}_{\mysw{R}i\alpha} (-i v_{i} \partial_{x}) \psi_{\mysw{R}i\alpha} + \psi^{\dag}_{\mysw{L}i\alpha} (i v_{i} \partial_{x}) \psi_{\mysw{L}i\alpha}.
\end{eqnarray}
The summation over the band index $i$ runs through all $N_{f}$ flavors. In general, the velocities are different and can not be eliminated by rescaling the space-time coordinates $(\tau, x)$ as in strictly one-dimensional systems.

Writing down all possible interactions is more laboring. By dimensional analysis, the most relevant interactions are the marginal four-fermion vertices. It turns out that the allowed vertices can be group together elegantly in terms of SU(2) currents. For convenience, we introduce the following SU(2) scalar and vector currents,
\begin{eqnarray}
J_{\mysw{P}ij} &=&\frac12\: \psi _{\mysw{P}i\alpha }^{\dag }\psi _{\mysw{P}j\alpha },
\\
\mybm{J}_{\mysw{P}ij}&=&\frac12\: \psi _{\mysw{P}i\alpha}^{\dag } \mybm{\sigma}_{\alpha \beta} \psi _{\mysw{P}j\beta },
\\
I_{\mysw{P}ij} &=& \frac12\: \psi _{\mysw{P}i\alpha }\epsilon_{\alpha \beta} 
\psi _{\mysw{P}j\beta},
\\
\mybm{I}_{\mysw{P}ij} &=& \frac12\: \psi _{\mysw{P}i\alpha }(\epsilon \mybm{\sigma})_{\alpha \beta}\psi _{\mysw{P}j\beta},
\end{eqnarray}
where $\epsilon_{12}=-\epsilon_{21}=1$ is the anti-symmetric Levi-Civita tensor and $\mybm{\sigma}$ are the Pauli matrices. The factor $1/2$ ensures the currents satisfy the conventional $SU(2)$ commutators.

At half filling, the particle-hole symmetry pairs up the Fermi momenta with the relation
\begin{eqnarray}
k_{\mysw{F}i} + k_{\mysw{F}\hat{\imath}} = \pi,
\label{PairKF}
\end{eqnarray}
where $\hat{\imath} = (N+1)-i$. Other than the above relation, different Fermi momenta are generally incommensurate. In loose terms, the Fermi surface (points, in fact) is only nested between the pairs of bands $i$ and $\hat{\imath}$, with the nesting condition $i+\hat{\imath}=N+1$. After some algebra, it is straightforward to show that all momentum-conserving vertices can be expressed in terms of products of the SU(2) currents,
\begin{eqnarray}
\mathcal{H}_{int}^{(1)} &=&
\tilde{c}_{ij}^{\rho} J_{\mysw{R}ij}J_{\mysw{L}ij}
-\tilde{c}_{ij}^{\sigma }\mybm{J}_{\mysw{R}ij} \cdot \mybm{J}_{\mysw{L}ij}
\nonumber\\  
&+&\tilde{f}_{ij}^{\rho} J_{\mysw{R}ii} J_{\mysw{L}jj}
-\tilde{f}_{ij}^{\sigma } \mybm{J}_{\mysw{R}ii} \cdot \mybm{J}_{\mysw{L}jj}  \nonumber \\
&+&\tilde{s}_{ij}^{\rho} J_{\mysw{R}ij} J_{\mysw{L}\hat{\jmath}\hat{\imath}}-\tilde{s}_{ij}^{\sigma }\mybm{J}_{\mysw{R}ij} \cdot 
\mybm{J}_{\mysw{L}\hat{\jmath}\hat{\imath}}.
\label{int1}
\end{eqnarray}
The couplings $\tilde{c}_{ij}$ and $\tilde{f}_{ij}$ denote the Cooper and forward scattering between bands $i$ and $j$. The strange vertex $\tilde{s}_{ij}$, involving four different bands, arises from the pairing up of Fermi momenta, as described in Eq.~\ref{PairKF}. The superscripts $\rho, \sigma$ label the charge and spin sectors of the couplings respectively.

Because $\tilde{f}_{ii},\tilde{c}_{ii}$ describe the same vertex, to avoid double counting, we choose the diagonal piece of the forward scattering amplitude to vanish, i.e. $\tilde{f}_{ii}=0$. The same reason leads to $\tilde s_{ii}=0=\tilde s_{i \hat{\imath}}$. While it is not obvious at this point, the choice of signs for the scalar and vector couplings in Eq.~\ref{int1}\ is such that they are all positive for the repulsive on-site interaction.

In addition to the vertices which conserve momentum exactly, there are also Umklapp interactions which conserve momentum only up to reciprocal lattice vector, $\Delta P = \pm 2\pi$,
\begin{widetext}
\begin{eqnarray}
\mathcal{H}_{int}^{(2)} &=&\frac{\tilde{u}_{ij}^{\rho}}{2}
(I_{\mysw{R}ij}^{\dag} I_{\mysw{L}\hat{\imath}\hat{\jmath}}
+I_{\mysw{L}\hat{\imath}\hat{\jmath}}^{\dag} I_{\mysw{R}ij})
-\frac{\tilde{u}_{ij}^{\sigma }}{2}
(\mybm{I}_{\mysw{R}ij}^{\dag} \cdot 
\mybm{I}_{\mysw{L}\hat{\imath}\hat{\jmath}}
+\mybm{I}_{\mysw{L}\hat{\imath}\hat{\jmath}}^{\dag }\cdot
\mybm{I}_{\mysw{R}ij})
\nonumber \\
&+&\frac{\tilde{w}_{ij}^{\rho }}{2}
(I_{\mysw{R}i\hat{\imath}}^{\dag }I_{\mysw{L}j\hat{\jmath}}
+I_{\mysw{L}j\hat{\jmath}}^{\dag }I_{\mysw{R}i\hat{\imath}})
-\frac{\tilde{w}_{ij}^{\sigma}}{2}
(\mybm{I}_{\mysw{R}i\hat{\imath}}^{\dag } \cdot \mybm{I}_{\mysw{L}j\hat{\jmath}}
+\mybm{I}_{\mysw{L}j\hat{\jmath}}^{\dag} \cdot 
\mybm{I}_{\mysw{R}i\hat{\imath}}).
\label{int2}
\end{eqnarray}
\end{widetext}
Again, since $\tilde{u}_{ii},\tilde{w}_{ii}$ describe the same vertex, we set
$\tilde{w}_{ii}=0=\tilde{w}_{i\hat{\imath}}$ to avoid double
counting. Note that both kinds of vertices involve four different bands in general.

Away from half filling, the relation in Eq.~\ref{PairKF} is no longer valid. As a result, the couplings $\tilde{s}_{ij}, \tilde{u}_{ij}, \tilde{w}_{ij}$ all become irrelevant because the fast oscillating phase arise from the finite momentum associated with the vertex. Therefore, we are left with the familiar Cooper and forward scattering\cite{Footnote1} in the doped $N$-leg ladder.

All vertices discussed in above share a unique feature. While the vertices in Eq.~\ref{int1} and \ref{int2} may involve four different bands, there are at most two different velocities associated with each vertex because $v_{i}=v_{\hat{\imath}}$, i.e. the velocities in each vertex always appear pairwise. This seemingly useless feature turns out to be strong enough to guarantee the existence of the RG potential when the Hamitonian is re-expressed in terms of Majorana fermions.

\section{Majorana Representation}

Now we switch to the Majorana fermion basis and construct the RG potential explicitly. Without interactions, the
band structure in low-energy limits is
described by $N_f$ flavors of Dirac fermions with {\em different} velocities in general. Each flavor of Dirac fermions can be decomposed into two Majorana fermions. Combined with spin degeneracy, the $4 N_{f}$ flavors of Majorana fermions are described by the Hamiltonian density
\begin{equation}
{\cal H}_{0} = \eta_{\mysw{R}a} (-i v_{a}\partial_x)
\eta_{\mysw{R}a} + \eta_{\mysw{L}a} (i
v_{a}\partial_x) \eta_{\mysw{L}a},
\end{equation}
where $v_a$ denotes the Fermi velocity for each flavor.

\begin{figure}
\centering
\includegraphics[width=5cm]{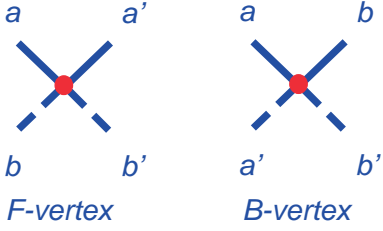}
\caption{\label{fig:Vertices} Forward and backward vertices. Bold
lines represent right-moving Majorana fermions while dashed lines
stand for the left-moving ones.}
\end{figure}

In general, a single vertex involves four different Fermi points, which generally would have four different velocities. However, the momentum conservation in weak coupling gives rise to the interesting constraint that the Fermi velocities must equal pairwise, as demonstrated explicitly in previous section. The interacting Hamiltonian in terms of the Majorana fermions take the form,
\begin{eqnarray}
{\cal H}_{int} &=& \tilde{F}(a,a';b,b') \eta^{}_{\mysw{R}a}
\eta^{}_{\mysw{R}a'} \eta^{}_{\mysw{L}b} \eta^{}_{\mysw{L}b'}
\nonumber\\
&+& \tilde{B}(a,b;a',b') \eta^{}_{\mysw{R}a} \eta^{}_{\mysw{R}b}
\eta^{}_{\mysw{L}a'} \eta^{}_{\mysw{L}b'},
\label{FBVertices}
\end{eqnarray}
where summations over allowed indices are implied. We emphasize that the allowed interactions might involve four different Fermi points labeled by $a,a',b,b'$, but only two different velocities $v_{a}=v_{a'}$ and $v_{b}=v_{b'}$ appear in a single vertex. By direct comparison, it is clear that the $F-$vertex include $\tilde{f}_{ij}$ and $\tilde{w}_{ij}$ and the $B-$vertex cover $\tilde{c}_{ij}$, $\tilde{s}_{ij}$ and $\tilde{u}_{ij}$.

There are also other kinds of interactions allowed by momentum
conservation,
\begin{eqnarray}
H_{c} &=&  \tilde{R}(a,a';b,b') \eta^{}_{\mysw{R}a} \eta^{}_{\mysw{R}a'}
\eta^{}_{\mysw{R}b} \eta^{}_{\mysw{R}b'} 
\nonumber\\
&+& \tilde{L}(a,a',b,b')
\eta^{}_{\mysw{L}a} \eta^{}_{\mysw{L}a'} \eta^{}_{\mysw{L}b}
\eta^{}_{\mysw{L}b'}. 
\label{ChiralVertices}
\end{eqnarray}
The scaling dimensions of these vertices are $(\Delta_{\mysw{R}},
\Delta_{\mysw{L}}) = (2,0), (0,2)$, while the vertices in
Eq.~(\ref{FBVertices}) have scaling dimensions $(1,1)$. Since the
renormalization comes from loop integrations, only vertices with
scaling dimensions differed by $(n, n)$, where $n$ is an integer,
would renormalize each other. As a result, the chiral vertices in
Eq.~(\ref{ChiralVertices}) remain marginal and only renormalize
the corresponding Fermi velocities. Since the corrections only
show up at two-loop order, we would ignore their contribution
here.  The pairwise-equal Fermi velocities in
Eq.~(\ref{FBVertices}) make the classification of all vertices
fairly simple as shown in Fig.~\ref{fig:Vertices}. The names come from the fact that the forward-type ($F$-) vertices include the usual forward scatterings while the backward-type ($B$-) vertices include the backward scatterings.

To obtain the flow equations, we need to integrate out
fluctuations at shorter length scale successively. The most
convenient approach is by the operator product expansions (OPE) of
these vertices which form a close algebra. The detail techniques to compute the renormalized interaction $\delta{\mathcal
  H}_{R}$ can be found in Ref.[\onlinecite{Lin97}].

\begin{figure}
\centering
\includegraphics[width=8cm]{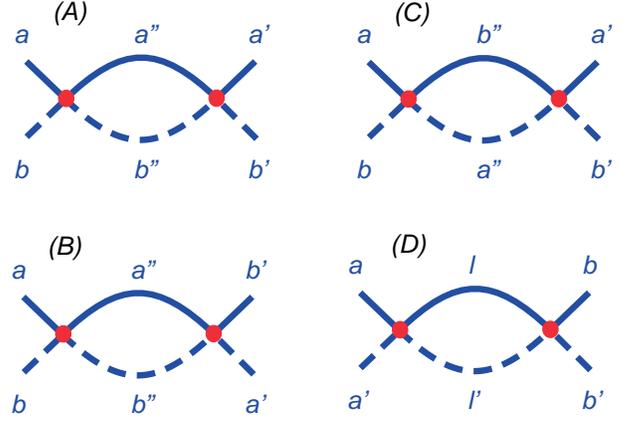}
\caption{\label{fig:Loops} Four different diagrams to the one-loop
order. (A) FF $\to$ F (B) FB $\to$ B (C)BB $\to$ F (D) BB $\to$ B.
Notice that, only in the fourth diagram, there are three
velocities involved while only two velocities are involved in all
other diagrams.}
\end{figure}

To one-loop order, the renormalization of the bare couplings come
from four types of diagrams $FF \to F$, $FB \to B$, $BB \to F$ and $BB \to
B$, shown in Fig.~\ref{fig:Loops}. Let us start with the first type of loop diagrams, $FF \to F$ in Fig.~\ref{fig:Loops}(a). The OPE of
Majorana fermions can be computed
straightforwardly and the mode elimination leads to the
renormalized Hamiltonian density,
\begin{eqnarray}
\delta {\cal H}^{R} &=& \tilde{F}(a,a'';b,b'') \tilde{F}(a'',a';b'',b')
\frac{\eta^{}_{\mysw{R}a} \eta^{}_{\mysw{R}a'}
\eta^{}_{\mysw{L}b} \eta^{}_{\mysw{L}b'}}{\pi (v_a+v_b)} dl
\nonumber\\
&=&  d\tilde{F}(a,a';b,b') \eta^{}_{\mysw{R}a} \eta^{}_{\mysw{R}a'}
\eta^{}_{\mysw{L}b} \eta^{}_{\mysw{L}b'},
\end{eqnarray}
where $dl = \ln b$ is the logarithmic length scale. Here we have used the fact that $v_{a}=v_{a''}=v_{a'}$ and $v_{b}=v_{b''}=v_{b'}$. The factor
$1/\pi(v_a+v_b)$ arises from the product of propagators with opposite chiralities and different velocities.

Introducing a simple rescaling of the original couplings according
to their associated velocities 
\begin{equation}
F(a,a';b,b') = \frac{1}{2\pi \sqrt{v_a v_b}}
\tilde{F}(a,a';b,b'),
\end{equation}
the RG equation is
\begin{equation}
\frac{dF(a,a';b,b')}{dl} = \gamma_{ab} F(a,a'';b,b'') F(a'',a';b'',b') + ...,
\end{equation}
where $ \gamma_{ab} = 2\sqrt{v_a v_b}/(v_a +v_b)$. Repeating similar calculations, the RG equation for $F(a,a'';b,b'')$ contains a term $\gamma_{ab} F(a,a';b,a') F(a'',a';b'',b)$ and similar result for the renormalization of $F(a'',a';b'',b')$. Therefore, the flow equations can be derived from a potential,
\begin{equation}
V(F) = - \gamma_{ab} F(a,a';b,b') F(a,a'';b,b'') F(a'',a';b''b').
\end{equation}

The RG potential for the $(FBB)$ and $(BBB)$ cases shown
in Fig.~\ref{fig:Loops}(b-c) and \ref{fig:Loops}(d) can be constructed in a similar fashion. Finally, combining all contributions together, the weak-coupling RG flows for Q1D systems are described by the potential
\begin{eqnarray}
\lefteqn{V(F,B) = -B(a,b;a',b') B(b,c;b',c') B(c,a;c',a')}
\nonumber\\
&-& \gamma_{ab} B(a,b';a',b) B(a'',b';a',b'') F(a,a'';b,b'')
\nonumber\\
&-&
\gamma_{ab} F(a,a';b,b') F(a,a'';b,b'') F(a'',a';b''b'),
\label{RGPotential}
\end{eqnarray}
Summations over all allowed indices are again implied. The merits to use Majorana representation enables us to construct the RG potential explicitly. Note that if one starts from a `wrong' basis, it is far from trivial to realize the fact that the non-linear flows can be derived from a single potential. However, in the Majorana representation, the linear transformation to the potential basis is diagonal $L_{ij} = r_{i} \delta_{ij}$. After appropriate rescaling of couplings, the explicit form of the RG potential is derived. We have checked for the doped and half-filled Q1D systems and found all potentials agree with Eq.~(\ref{RGPotential}).

There is one loose end about the rescaling factors. For most physical systems, the rescaling factor is slightly more complicated than $(2\pi \sqrt{v_a v_b})^{-1}$. the subtlety arises from the degeneracies of the couplings imposed by physical symmetries. This is best illustrated by the following simple example. Consider the RG equations for three couplings $g_i$,
where $i=1,2,3$,
\begin{equation}
\frac{dg_{i}}{dl} = \sum_{jk} \frac{|\epsilon_{ijk}|}{2} g_{j} g_{k}.
\end{equation}
Since $|\epsilon_{ijk}|$ is totally symmetric, the corresponding
RG potential is $V(g) = g_1 g_2 g_3$. Suppose the system has some
symmetry, such as $U(1)$ symmetry for charge conservation, and the
couplings are degenerate $g_2 =g_3$. The RG equations are simplified,
\begin{equation}
\frac{dg_{1}}{dl} = g_2^2, \qquad \frac{dg_{2}}{dl} = g_1 g_2.
\end{equation}
It is straightforward to show that we need to perform a rescaling
transformation, $(h_1, h_2) = (g_1, \sqrt{2} g_2)$ to obtain the
potential $V(h) = h_1 (h_2)^2/2$. In fact, for couplings with
$n$-fold degeneracy, an additional rescaling factor $\sqrt{n}$ is
necessary to bring them into the potential basis. Therefore, the total rescaling factor is
\begin{equation}
r_{i} = \frac{1}{2\pi} \sqrt{\frac{n_i}{v_a v_b}},
\end{equation}
where $n_i$ is the degeneracy number of the coupling $g_i$, with Fermi velocities $v_a$ and $v_b$.

\section{Conclusions and Discussions}

So far, we have shown the existence of the RG potential by explicit construction, which proves the the widely used Ansatz in Eq.~\ref{Scaling}. In fact, the asymptotes of the RG flows are governed by the special set of fixed-ray solutions. This in turns explains the simplicity of the phase diagram, even though the RG equations are rather complicated.

The absence of exotic fates of the RG flows is also found in earlier work on 2D melting theory\cite{Kosterlitz73} or the Kondo related problems\cite{Anderson70}. However, the simplicity of RG flows in these systems is not quite the same as described here. Since the number of marginal couplings in these problems is few, analytic solution of the flows often show that it is possible to define some conserved quantity associated with each flow lines. This is where the simplicity comes from. On the other hand, we have also looked into these well-known flows to check whether they can be derived from a single potential. It is not too surprising that this is indeed the case because the requirement of potential flows loose up quite a bit when the number of couplings is small.

With the help of non-perturbative Abelian bosonization, it is not essentially important whether the RG flows can be cast into potential form beyond one-loop order. However, it remains an interesting and open question at this moment. Note that the coefficients of the one-loop RG equations are unique, protected by the leading logarithmic divergences. The next order calculations bring in lots of complications and subtleties, including the non-universal coefficients in the RG equations, velocity renormalization and so on. It is not clear at this moment whether it is even sensible to pursue the RG potential beyond one-loop order.

Another interesting issue concerns the connection between the
potential $V(h_i)$ and the Zamolodchkov's c-function $C(g_i)$ of
(1+1)-dimensional systems with Lorentz and translational
symmetries\cite{Zamolodchikov86}.  A generic Q1D system we studied here has neither Lorentz invariance (due to different Fermi
velocities) nor translation symmetry (due to Umklapp processes).
While both $V$ and $C$ are non-increasing along the RG flows, the exact
relation between them remains unclear at this point.  We emphasize
that the existence of a non-decreasing function $C$ along RG flows
only implies that $dC/dl = (\partial C/\partial g_i)\cdot
(dg_i/dl) \leq 0$ and is not strong enough to show that the flows
can be derived from a potential.  Thus, the potential flows are
closely related to the c-theorem but they are not equivalent in
general. In addition, we do not know any easy generalization of
c-theorem that does not rely on the Lorentz and translational
symmetries.  However, one can easily check that in the special
limiting case where Lorentz and translation
symmetries are restored, the $C$-function indeed coincides with the potential we
find.  This indicates that there may be a general form of
c-theorem waiting to be discovered.

In conclusion, we have shown that the RG transformation for Q1D
systems in weak coupling is described by potential flows.
Therefore, neither chaotic behaviors nor exotic limit cycles could
occur. The different Fermi velocities and the degeneracies
imposed by physical symmetries give rise to non-trivial rescaling
factors, which hinder this beautiful structure behind the RG
transformation. The explicit form of the potential is obtained
after appropriate rescaling of the couplings in Majorana basis.

We thank Leon Balents, Matthew Fisher, Y.-C. Kao, Andreas Ludwig and S.-K. Yip for fruitful discussions. In particular, HHL is thankful for illuminating counter example in dynamical systems provided by Leon Balents. HHL appreciates financial supports from National Science Council in Taiwan through Ta-You Wu Fellow and grants NSC-91-2120-M-007-001 and NSC-92-2112-M-007-039. CYM acknowledges support from NSC of Taiwan under grant NSC-92-2112-M-007. The hospitality of KITP, where part of the work was carried out, and supports from NSF PHY99-07949 are great appreciated.

\end{document}